%% file: paper.tex
\documentstyle{europhys}
\input euromacr

\input epsf.sty
\begin{document}
\euro{}{}{}{}
\Date{}
\shorttitle{X. WAINTAL \etal Delocalized Coulomb phase in $2d$}
\title{Delocalized Coulomb phase in two dimensions}
\author{Xavier Waintal, Giuliano Benenti \And Jean-Louis Pichard}
\institute{
CEA, Service de Physique de l'Etat Condens\'e, \\
Centre d'Etudes de Saclay, F-91191 Gif-sur-Yvette cedex, France   
}
\rec{}{}
\pacs{
\Pacs{71}{30+h} {Metal-insulator transitions and other electronic transitions}
\Pacs{72}{15Rn} {Quantum localization}
\Pacs{71}{27+a} {Strongly correlated electron systems} 
}
\maketitle
\begin{abstract}
 Extending finite size scaling theory to the many body ground state, 
 one finds that Coulomb repulsion can drive a system of spinless 
 fermions in a random potential from the Anderson insulator (Fermi 
 glass of localized states) towards a new extended phase  
 in dimension $d=2$. The transition occurs at a Coulomb energy to Fermi 
 energy ratio $r_s \approx 4$, where a change in the nature of the 
 persistent currents has been previously observed. Relevance to the 
 recently observed $2d$ metallic phase is suggested. 

\end{abstract}
%
%

 Following the scaling theory of localization \cite{aalr}, all $2d$ systems 
of electrons (or holes) are localized when electron-electron interaction 
is negligible. This absence of metals in two dimensions is nowadays 
questioned after the observation of a transition towards a metallic behavior 
in silicon based $2d$ electron gases \cite{kravchenko,popovic,pudalov} 
when the carrier density is varied. This unexpected phenomenon 
occurs also for hole gases in GaAs heterostructures \cite{hanein,hamilton}, 
SiGe \cite{coleridge} and InAs quantum wells \cite{ensslin}. The transition 
is observed at very low carrier concentrations, the 
charge spacing being $\geq 10^3 \AA $ in certain cases \cite{mills}. This 
suggests that Coulomb repulsion is the driving mechanism for the transition. 
In ref. \cite{yoon}, a very clean heterostructure was studied, and the 
transition was observed at $r_s \approx 35$, close to $r_s \approx 37$ 
for which Wigner crystallization is expected without disorder. In more 
disordered samples, 
the transition occurs typically around $r_s \approx 10$ and $k_F l 
\approx 1$, $k_F$ denoting the Fermi wave vector and $l$ the elastic mean free 
path. It is also at $r_s \approx 10$ that charge crystallization 
is numerically observed in small disordered clusters \cite{tanatar,bwp1}. 
This gives arguments to associate the observed transition to the quantum 
melting of a kind of pinned Wigner crystal. In this case, the metallic 
phase should cease to exist at a weaker $r_s$ also and the re-entry 
towards a Fermi glass of weakly interacting localized particles should 
occur. The re-entry has been observed in Ref. \cite{hamilton} at  
$r_s \approx 6$ (see also Ref. \cite{pudalov}). 
Those regimes are out of reach 
of the Landau theory of disordered metals \cite{aa,finkelstein} valid for 
large $k_Fl$  and small $r_s$. 

The interplay between disorder and electron-electron interactions 
is attracting increasing theoretical interest \cite{hamburg}, 
though many studies are devoted to large energy excitations \cite{cuevas} 
of a few particle states instead of the ground state properties. In 
Ref \cite{bwp1}, it was shown that the ground state of a system of spinless 
fermions goes from an insulating phase (Fermi glass, $r_s \leq r_s^F$) towards another 
insulating phase (pinned Wigner crystal, $r_s \geq r_s^W$) through 
a new intermediate quantum phase. This conclusion was drawn from a study 
of the persistent currents supported by the ground state in small 
clusters. A $2d$ torus geometry enclosing a flux $\phi$ around the longitudinal 
direction was considered. The ratio $r_s^F$ characterizes the suppression of the 
transverse current $I_t$ (not enclosing $\phi$). Below $r_s^F$, the pattern 
of driven currents has a $2d$ topology and the longitudinal current $I_l$ 
(enclosing $\phi$) can be paramagnetic or diamagnetic, depending on 
the sample. Above $r_s^F$, the flow pattern has an ordered $1d$ 
topology \cite{bwp1,avishai} 
and the sign of $I_l$ becomes sample independent. The higher ratio $r_s^W$ 
characterizes the suppression of $I_l$ and charge crystallization in a random 
substrate. For $0.3 < k_Fl < 3$, one obtains $r_s^W \approx 10$ and 
$r_s^F \approx 4$, in agreement with ratios $r_s \approx 10$ where the 
metal-insulator transition is observed, and with a ratio $r_s \approx 6$ 
where the re-entry has been reported in Ref. \cite{hamilton}. Though small 
clusters exhibit an enhancement of $I_l$ for intermediate $r_s$, exact 
diagonalization does not allow to vary system size and to establish that 
the intermediate phase is metallic at the thermodynamic limit. To understand 
how the cooperative Coulomb behavior begins to destroy Anderson localization 
of weakly interacting particles, and if it drives the system towards a 
delocalized phase before yielding charge crystallization, we need to 
extend the finite size scaling method \cite{ps1,ps2,kmk1,kmk2} successful for 
describing single particle delocalization to the many body ground state. 
We do not study in this work large Coulomb repulsions where charge crystallization 
sets in, but using approximations valid for weaker repulsions, one finds that the Fermi 
glass melts at $r_s \approx r_s^F \approx 4$ (in good agreement with Ref. \cite{bwp1}) 
to give rise to a more fluid phase for the particles.  
Using a low carrier density, we characterize this 
melting by a suitable localization length which displays in $d=2$ a scaling 
behavior similar to the one body (1B) localization length in $d=3$ at the 
mobility edge, i.e. more generally a behavior characteristic of second order 
phase transitions. This confirms that spinless fermions have an intermediate 
delocalized phase between the Fermi glass and the pinned Wigner crystal. 

  Anderson 1B localization is a complex phenomenon which is analytically 
tractable in a few limits: quasi-one dimension \cite{larkin,pichard} and 
Bethe lattice \cite{aat}. For dimensions $d=2$ and $3$, our knowledge 
of 1B localization is mainly based on numerical works \cite{ps1,ps2,kmk1,kmk2}. 
The successful method consists in evaluating the 1B localization length 
$L_1(L)$ of a system of finite size $L$ and to verify a scaling ansatz 
inspired \cite{ps1} from the theory of second order phase transitions: 
\begin{equation} 
\label{ansatz}
\frac{L_1 (L)}{L} = F_d \left( \frac{L}{L_1(d)} \right).
\end{equation} 
The lengths $L_1(L)$ calculated for different system parameters can be 
mapped onto a universal curve $F_d$ assuming a single scaling length 
$L_1(d)$ which characterizes the $d$-dimensional lattice. For the 1B 
problem, it was convenient to consider $d$-dimensional strips of finite 
transverse section $L^{d-1}$. $L_1(L)$ was defined as the inverse of the 
smallest positive Lyapunov exponent of the appropriate product of transfer 
matrices, which is a self averaging quantity. The ansatz (\ref{ansatz}) 
was verified in dimensions $d \geq 1$, and a metal-insulator transition 
was obtained \cite{ps1,ps2,kmk1,kmk2} for $d=3$ with $L_1(3d)$ diverging 
at the mobility edge. In two dimensions, large $L$ studies \cite{kmk1} led 
to the conclusion that $L_1(2d)$ diverges only in the clean limit, in 
agreement with Ref.\cite{aalr}. 

  We extend this finite size scaling method to the ground state of 
$N$ spinless fermions with Coulomb repulsion in a random potential 
defined on a square lattice with $L^2$ sites. 
The Hamiltonian reads: 
\begin{eqnarray} 
\label{hamiltonian} 
H=-t\sum_{<i,j>} c^{\dagger}_i c_j +  
\sum_i v_i n_i +U \sum_{i\neq j} \frac{n_i n_j}{2 r_{ij}}. 
\end{eqnarray} 
$c^{\dagger}_i$ ($c_i$) creates (destroys) a particle in 
the site $i=(i_x,i_y)$, $t$ is the strength of the hopping terms 
between nearest neighbors (kinetic energy) and $r_{ij}$ 
is the inter-particle distance for a $2d$ torus (periodic 
boundary conditions). The random potential $v_i$ of the site 
$i$ with occupation number $n_i=c^{\dagger}_i c_i$ is taken 
from a box distribution of width $W$. The interaction strength 
$U$ yields a Coulomb energy to Fermi energy ratio 
$r_s=U/(2t\sqrt{\pi n_e})$ for a filling factor $n_e=N/L^2$. 

 To investigate the melting of the Fermi glass by Coulomb repulsions,  
one begins with a low density of occupied 1B states such that the 
charge spacing is larger than $L_1$ without interaction. The size $L$ 
varies from $L=24$ ($N=3$) to $L=31$ ($N=5$), $N$ being chosen 
to keep $n_e=1/192$. An intermediate size $L=28$ ($N=4$) with almost 
the same density ($1/196$) is also considered.

\begin{figure}
\centerline{ 
\epsfxsize=14cm
\epsfbox{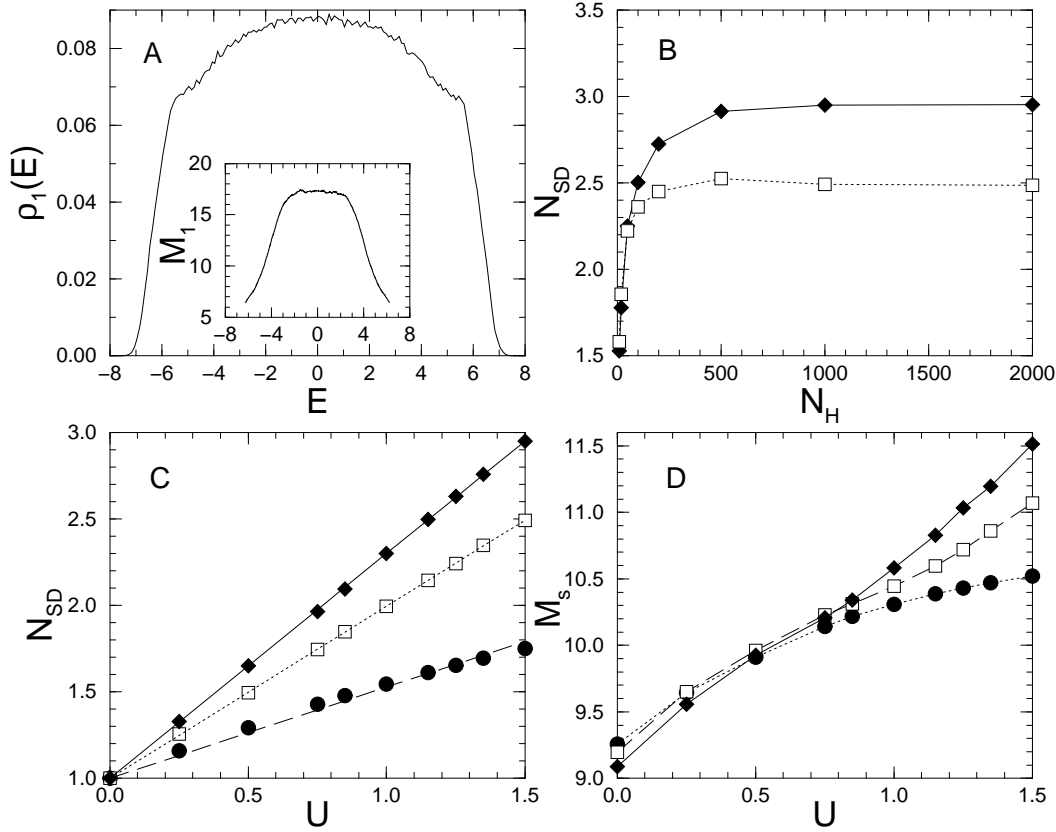}
}
\caption{
{\bf A}: 1B density of states $\rho_1(E)$ and number $M_1 \approx 
L_1^2$ of occupied sites (insert) as a function of the energy for 
$W=10$ and $L=31$. {\bf B}: Convergence of the number $N_{SD}$ of 
SD occupied by the ground state when $N_H$ increases. $L=31$ (upper curve) 
and $L=28$ (lower curve) at $U=1.5$. {\bf C}: Variation of $N_{SD}$ as 
a function of $U$ for $L=24$ (circle), $28$ (square) and $31$ (diamond). 
{\bf D}: Number $M_s$ of occupied sites per particle as a function of $U$ 
for the three values of $L$ (same symbols than beside). 
}
\label{fig1}
\end{figure}

 Let us first discuss some features of the 1B spectrum. If we take a 
small value for $W$, strong lattice effects remain for $L\approx 31$, 
complicating the extrapolation towards the thermodynamic limit. The lattice 
effects disappear at larger $W$, but another difficulty remains, due to the 
low density $n_e$ numerically accessible. The low energy tail of the 
1B spectrum is made of impurity states where the particles are simply 
trapped at some site $i$ of exceptionally low $v_i$. Their energies are 
below the band $[-4t, 4t]$ of the clean system. This is a trivial 
localization with $L_1 \leq 1$ and irregular energy spacings 
given by the tails of the distribution of the $v_i$. We are not interested 
to study the detrapping of those impurity states by Coulomb repulsions, 
but by the delocalization of genuine Anderson localized states, of energy 
$E>-4t$. The localization of those states results from the multiple 
scattering of plane waves, while the impurity states involve mainly 
a few evanescent waves. With a higher density, it would be easy to fill 
the tail of 1B band and to put the Fermi energy inside the band of 
Anderson levels of energy $>-4t$, having regular energy spacings 
$\Delta_1 \propto 1/L^2$ and similar $L_1$ (see Fig. \ref{fig1} A). With 
a low density, in order to observe nevertheless a delocalization resulting 
from the mixing by the interaction of SD built out from Anderson localized 
states (say with $L_1 \approx 4$), we calculate the $L^2$ 1B levels for 
$W=10$ and we ignore the $L^2/2$ first 1B levels, considering only as 
possible 1B states the $L^2/2$ remaining ones. Doing so, the states near 
the Fermi energy of the non interacting system are not simple impurity states. 
We have checked that all the results presented hereafter are identical if we 
get rid of a smaller fraction of 1B states, as far as the Fermi energy at $U=0$ 
is located in the constant part of the 1B density of states (Fig.\ref{fig1} A).
For $W=10$ and 1B states around the 1B band center, one can roughly estimate that 
$k_Fl \approx 1$. From this restricted subset of 1B Anderson localized states, 
the $N_H$ first SD ordered by increasing energy are calculated. We assume that 
such a procedure captures genuine $2d$ ground state physics, since the 
low energy excitations begin with a (large) 1B energy spacing $\Delta_1$, 
and not with a (very small) NB energy spacing $\Delta_N \propto L^{-2N}$.

 The sizes ${\cal N}_H=(L^2)!/(N!(L^2-N)!)$ of the corresponding Hilbert 
space are huge and only the projection of the Hamiltonian ($\ref{hamiltonian}$) onto 
smaller subspaces can be diagonalized (Lanczos method). For weak $U$, we consider 
the subspace spanned by the $N_H$ Slater determinants (SD) corresponding 
to the first low energy states of the non interacting problem. The $N$ body 
(NB) ground state $|\Psi_0>$ is obtained after diagonalizing the truncated 
Hamiltonian. $|\Psi_0>= \sum_i C_i |SD_i>$ is typically extended in the SD basis 
over $N_{SD} = <\sum_i |C_i|^4>^{-1}$ SD, the brackets denoting the average 
over $5 \times 10^3$ disordered samples. In Figs. \ref{fig1} B-C, one can see 
that $N_{SD}$ remains stable when $ N_H \geq 500$ at a value negligible 
compared to $N_H=1000$  (value for $N_H$ assumed hereafter). 
Though the stability of our results have been checked when $N_H$ varies 
inside a range negligible compared to ${\cal N}_H$, making difficult to rule 
out the existence of a slow variation which cannot be detected 
inside a too narrow range, we assume that $|\Psi_0>$ has a negligible chance 
to have a significant projection outside the $1000$ first SD 
when $L \leq 31$ and $U \leq 1.5$. 

 Figure \ref{fig1} D gives the number $M_s = N <\sum_i \rho_i^2>^{-1}$ 
of occupied sites per particle in the ground state, 
$\rho_i = <\Psi_0 |n_i |\Psi_0>$ denoting the charge density of the 
ground state at site $i$. For $U=0$, $M_s$ is obviously 
smaller than $L_1^2$, a limit where the $N$ occupied 1B states do not 
overlap. When one goes towards the thermodynamic limit, 
$M_s$ is almost size independent for the Fermi glass ($U\leq 1$), but 
varies as a function of $L$ when $U \geq 1$. This provides a first 
evidence that one has a delocalization effect at an interaction threshold 
$U \approx 1$. The glassy ground state melts towards a more fluid phase 
which fills a larger fraction of the sample. This observed delocalization  
cannot be related to charge crystallization, since $M_s$ increases while 
$M_s \rightarrow 1$ when $U \rightarrow \infty$.

\begin{figure}
\centerline{
\epsfxsize=14cm
\epsfbox{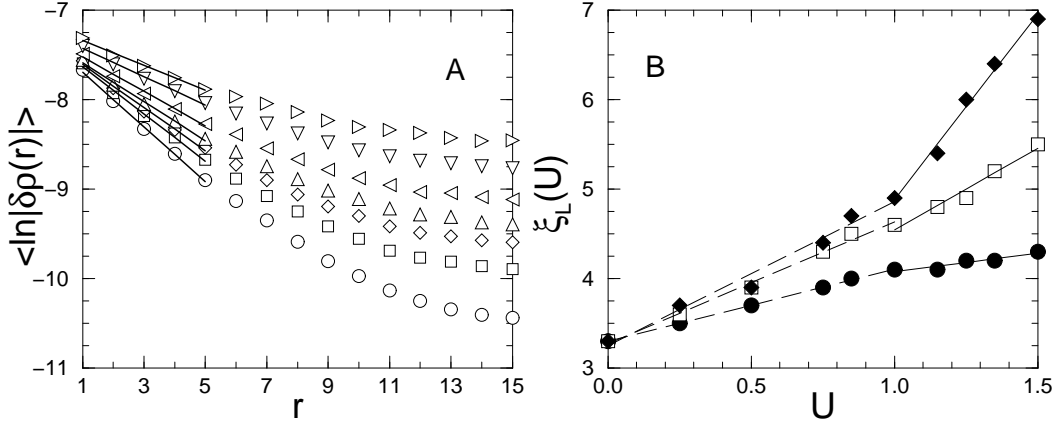}
}
\caption{
{\bf A}: $<\ln |\delta \rho(r)|>$ as a function of $r$ for $L=31$ at 
$U=0$ (circle), $0.25$ (square), $0.5$ (diamond), $0.75$ (triangle 
up), $1$ (triangle left), $1.25$ (triangle down) and $1.5$ (triangle 
right). The lengths $\xi_L(U)$ are given by the slopes of the thick lines. 
{\bf B}: $\xi_L$ as a function of $U$ for $L=24$ (circle), $28$ (square) 
and $31$ (diamond).
}
\label{fig2}
\end{figure}
 
 To characterize the NB ground state by an appropriate localization 
length, we consider the change $\delta \rho_j$ of the charge density 
induced by a small change $\delta v_i$ of the random potential $v_i$ located 
at a distance $r=|i-j|$. For a Slater determinant made with N occupied 1B 
eigenfunctions ( $\psi_\alpha$), first order perturbation theory gives: 
\begin{equation} 
\delta\rho_j = 2\delta v_i\sum_{\alpha=1}^N\sum_{\beta\neq \alpha}
\frac{\psi_\alpha(i)\psi_\beta(i)\psi_\alpha(j)\psi_\beta(j)}{E_\beta
-E_\alpha}\propto\exp-\frac{2r}{L_1},
\end{equation} 
the index $\beta$ varying over the 1B spectrum. Therefore, the change 
$\delta \rho$ remains localized on a scale $\xi_L \approx L_1/2$ without 
interaction. We study how $\xi_L$ depends on 
$U$ for different values of $L$. To improve the statistical convergence, 
we calculate more precisely the change $\delta \rho (r) = 
\sum_{j_y} \delta \rho_{r,j_y}$ of the charge density on the $L$ 
sites of coordinate $j_x=r$ yielded by the change $v_{0,i_y} 
\rightarrow v_{0, i_y} (1+1/100)$ for the $L$ random potentials 
of coordinate $i_x=0$. The quantity $\delta \rho(r)$ for $W=10$ 
has a broad symmetric distribution of amplitude $|\delta \rho (r)|$ 
which is almost log-normal for $U=0$. There is no underlying law 
of large numbers which tells us what is the right self-averaging scaling 
variable, as Oseledec's theorem for the quasi-$1d$ 1B problem \cite{ps1}. 
Though the log-normal character of the distribution becomes less pronounced 
when $U$ increases, it still makes sense to characterize the typical strength 
of the fluctuations by
\begin{equation}
|\delta \rho (r)|_{\rm typ} = \exp <\ln | \delta \rho (r)| > 
\propto \exp-\frac {r}{\xi_L}, 
\end{equation}
where the brackets denote the average over $5 \times 10^3$ disordered samples. 
We extract the range $\xi_L$ of the perturbation from the $r$ dependence 
of this typical value inside a square of size $L$.

 An exponential decay of $|\delta \rho (r)|_{\rm typ}$ occurs only over a scale 
$ r << L/2$ since the boundary conditions are periodic. In Fig. \ref{fig2} A, 
one can see how the length $\xi_L$ are obtained from the slope of the linear parts 
of the curves (thick straight lines). Fig. \ref{fig2} B gives how 
$\xi_L$ depends on $U$ for the three considered sizes. This figure conveys 
a very different information than Fig. \ref{fig1} C. The number 
$N_{SD}$ of SD over which the NB ground state has important projections linearly 
increases as a function of $U$, without exhibiting a change of behavior. This rules out 
that $U$ can induce a sharp localized-delocalized transition in Hilbert space of 
the type suggested in Ref. \cite{agkl} for metallic quantum dots. The behavior 
of $\xi_L (U)$ is much more interesting. When $L=31$ and $U \geq 1$, the 
range of a local perturbation has a sharper increase which can only be due to 
the nature of the last SD participating to the NB ground state, and not to their 
number $N_{SD}$. The last SD participating to the ground state when $U \geq 1$ 
should not be made of 1B states localized in the vicinity of those playing a role 
when $U \leq 1$. This may indicate that the underlying delocalization mechanism 
is related to variable range hopping between localized 1B states yielded 
by Coulomb repulsions. This possibility was discussed in 
Refs. \cite{fleishmann,anderson,pollak}.

\begin{figure}
\centerline{  
\epsfxsize=14cm
\epsfbox{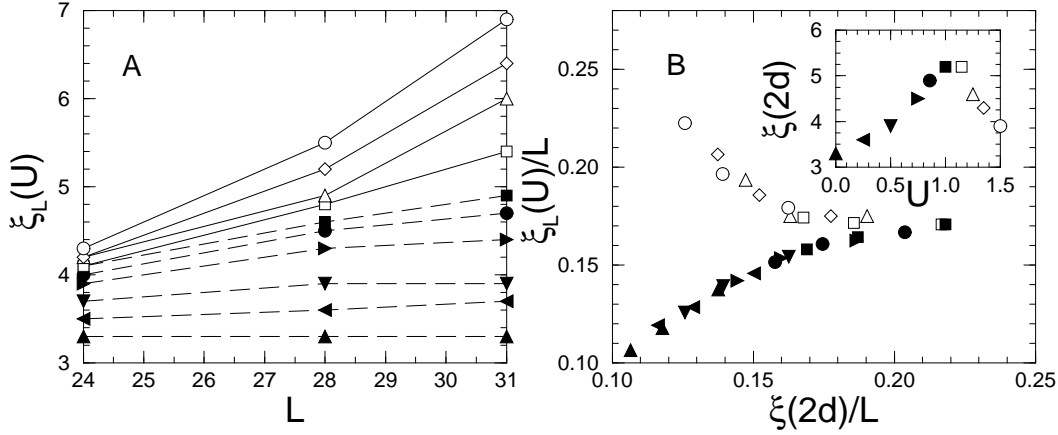}
}
\caption{ 
{\bf A}: Localization lengths $\xi_L(U)$ as a function of $L$ for 
$U \leq 1$ (filled symbols): $0$ (triangle up), $0.25$ (triangle left), 
$0.5$ (triangle down), $0.75$ (triangle right), $0.85$ (circle) 
and $1$ (square) and $U >1$ (empty symbols): $1.15$ (square), $1.25$ 
(triangle), $1.35$ (diamond) and $1.5$ (circle).
{\bf B}: Ratios $\xi_L(U) /L $ mapped onto the scaling curve $F_2$ 
as a function of $\xi(2d)/L$. The two dimensional scaling length 
$\xi (2d)$ is given in the insert.  
}
\label{fig3}
\end{figure}

The scaling analysis is shown in Fig. \ref{fig3}. The size dependence of 
$\xi_L$ is presented in Fig. \ref{fig3} A for increasing Coulomb 
repulsions. One finds the behavior typical of a transition: $\xi_L$ converges 
towards a finite value when $U <U_c$ (localized Fermi glass); diverges linearly 
as a function of $L$ at $U=U_c\approx 1$ (critical point) and diverges faster 
than linearly when $U>U_c$ (extended phase). This is exactly the 
behavior \cite{ps1} which characterizes the Anderson transition at $d=3$ for 
the 1B spectrum. In Fig. \ref{fig3} B, one verifies the scaling ansatz (\ref{ansatz}), 
where $\xi_L$ and $\xi(2d)$ play the role of $L_1(L)$ and $L_1(d)$ respectively. 
All the data of Fig. \ref{fig3} A can be mapped onto a universal curve $F_2$ 
shown in Fig. \ref{fig3} B, assuming the scaling length $\xi(2d)$ given in the 
 insert. When $U<U_c$, this length characterizes the localization 
of the effect of a local perturbation of the substrate in the two dimensional 
thermodynamic limit. Considering that the delocalization threshold is 
located at $U_c \approx 1$, 
a power fit of $\xi(2d) \propto |U-U_c|^{-\nu}$ yields a rough estimate for the 
critical exponent $\nu \approx 4$. More detailed and accurate studies of the 
vicinity of the delocalization threshold are necessary for confirming this 
value for $\nu$. Very often, additional corrections 
$\propto L^{-\alpha}$ to the scaling ansatz occurs for small sizes. 
We point out that our results can be fitted by a simple linear law $\xi_L =  
0.17 L$ for $U=U_c$. This is a further indication that the simple ansatz 
(\ref{ansatz}) describes scaling for $L \geq 24$, without 
noticeable additional $L^{-\alpha}$ corrections.  The obtained interaction 
threshold $U_c \approx 1$ gives $r_s \approx 4$, which agrees with 
the $r_s^F$ given in Ref. \cite{bwp1} for similar values of $k_Fl$ and is 
consistent with the threshold $r_s \approx 6$ where the re-entry is reported 
in Ref. \cite{hamilton}.

  In summary, we have shown that there is a melting of the Fermi glass 
towards a new delocalized phase at $r_s=r_s^F \approx 4$ and $k_Fl \approx 1$. 
In the glassy state, the particles occupy a tiny fraction 
$\leq n_e L_1^2$ of the sample and the effect of a local perturbation remains 
localized. Above $r_s^F$, the ground state becomes extended  
over the whole sample, as shown by the divergence of the range $\xi (2d)$ 
characterizing the two dimensional thermodynamic limit when $r_s \rightarrow r_s^F$. 
The range $\xi_L$ calculated in a square of size $L$ satisfies a finite size 
scaling ansatz consistent with a second order quantum phase transition 
and a power law divergence of $\xi(2d)$ at $r_s^F$. The effect of a local perturbation (motion of a 
single atom) have interesting implications for $1/f$ noise in the metallic 
\cite{feng} and localized \cite{fpz} regimes. Therefore, the implications of 
our numerical results might be checked by suitable noise measurements in 
low density $2d$ gases of electrons or holes.

This work is partially supported by a TMR network of the EU and by a 
Franco-Israeli AFIRST grant.

\end{document}

%% file: euromacr.tex
\def\etal{{\hbox{{\tenit\ et al.\/}\tenrm :\ }}}

\def\And{{\rm and\ }}

\newif\ifboo \boofalse

\def\Review#1{\boofalse{\it #1},}
\def\Name#1{{\sc #1},}
\def\Vol#1{\ifboo Vol. {\bf #1}\else{\bf #1}\fi}
\def\Year#1{\ifboo #1\else(#1)\fi}
\def\Book#1{\bootrue{\it #1},}
\def\Page#1{\ifboo {\rm p. #1}\else{\rm #1}\fi}